\documentclass[journal]{IEEEtran}
\ifCLASSINFOpdf

\else

\fi

\usepackage{color}
\usepackage{amsfonts}
\usepackage{amsmath}
\usepackage{graphicx}
\usepackage{amssymb}
\usepackage{stfloats}
\usepackage{algorithmicx,algorithm}
\begin{document}
\title{A Robust Secure Hybrid Analog and Digital Receive Beamforming Scheme for Efficient Interference Reduction}

\author{Linlin Sun, Yaolu Qin, Feng Shu, Riqing Chen, Yijin Zhang, Shuping Yang,\\ Jinhui Lu, Jun Li, and Jiangzhou Wang
\thanks{F. Shu and R. Chen are with the College of Computer and Information Sciences, Fujian Agriculture and Forestry University, Fuzhou 350002, China.
L. Sun, Y. Qin, F. Shu, Y. Zhang, S. Yang, J. Lu and J. Li are with School of Electronic and Optical Engineering, Nanjing University of Science and Technology, Nanjing, 210094, China.
J. Wang is with the School of Engineering and Digital Arts, University of Kent, Canterbury CT2 7NT, U.K.}}
\maketitle

\begin{abstract}
Medium-scale or large-scale receive antenna array with digital beamforming can be employed at receiver to make a significant interference reduction, but leads to expensive cost and high complexity of the RF-chain circuit. To deal with this issue, a classic analog-and-digital beamforming (ADB) structure
was proposed in the literature for greatly reducing the number of RF-chains. Based on the ADB structure, we in this paper propose a robust hybrid ADB scheme to resist directions of arrival (DOAs)
estimation errors. The key idea of our scheme is to employ null space projection (NSP) in analog beamforming domain and diagonal loading (DL) method in digital beamforming domain. Simulation results show that the proposed scheme performs more robustly, and moreover, has a significant improvement on the
receive signal to interference plus noise ratio (SINR) compared to NSP ADB scheme and DL method.
\end{abstract}

\section{Introduction}
Physical layer security problem has attracted more and more research interests from both academia and industry \cite{N.Zhao}, and now increasingly becomes one of the most important problems in wireless networks. In general, interference reduction \cite{J.Ma} is crucial in ensuring the secure acception of information, especially when enemies send interference signals to interrupt the reception of the desired signal \cite{J.Guo}. As a classic secure transmit method, directional modulation (DM) \cite{F.Shu1,F.Shu2} preserves the original signal constellation of transmitted signals along the desired direction well, and distorts the signal constellation along the undesired direction \cite{J.Hu}. For such a method, it is clear that the estimation errors of the directions of arrival (DOAs) measurement of the desired and undesired signals will lead to the performance loss in interference reduction.

To alleviate this negative impact, several robust interference reduction algorithms have been proposed in the literature to resist DOAs estimation errors. By reducing the mean square error (MSE), \cite{A.Hakam} presented a variable step size normalized least mean square (LMS) algorithm. \cite{A.M.J} proposed a self-adaptive algorithm for a robust direct-sequence spread-spectrum system, which is fast, loop structures avoiding, easy and cheap to implement in hardware. In \cite{Z.Li}, with the prior knowledge of possible target region, a robust deceptive interference reduction method based on covariance matrix reconstruction with frequency diverse array (FDA) multiple-input multiple-output (MIMO) radar is proposed. However, all of these schemes are based on the digital beamforming structure where each antenna needs one single radio frequency (RF) chain, so that the large array size and large hardware cost are inevitable.

In order to make a good balance between circuit cost and interference reduction performance simultaneously, we in this paper apply a hybrid analog-and-digital beamforming (ADB) structure into the receiver.
Two distinctive features of this structure are that each subarray output is viewed as one single virtual large antenna output and a digital beamforming operation is performed with an analog beamforming operation together.
Furthermore, built on this structure, we propose a robust hybrid ADB algorithm that jointly employs null space projection (NSP)~\cite{Y.Ding} and diagonal loading (DL) algorithm~\cite{Carlson}.
It will be shown by simulation that the proposed scheme is more robust to the DOA measurement errors and interference in comparison with existing NSP and DL methods.

It is known that the hybrid structure has been studied in many papers. In \cite{X.Zhang}, two hybrid structures, i.e., the fully-connected and partially-connected structures, were proposed to design a hybrid analog and digital precoding algorithm that can reduce the cost of RF chains. \cite{X.Yu} developed a low-complexity alternating minimization precoder by enforcing an orthogonal constraint on the digital precoder. In \cite{Alluhaibi}, two precoders based on the principle of manifold optimisation and particle swarm optimisation were proposed. An energy-efficient hybrid precoding for partially-connected architecture was proposed in \cite{X.Gao}. \cite{Ying} presented achievable rates of hybrid precoding in multi-user multiple-input multiple-output (MU-MIMO) system when employing only one RF chain per user and investigated the impact of phase error on hybrid structure performance. To make the optimal tradeoff between energy efficiency and spectrum efficiency, \cite{S.Han} achieved the green point for fixed product of the number of transceivers and the number of active antennas per transceiver. Meanwhile, \cite{A.Alkhateeb} proposed an iterative hybrid beamforming algorithm for the single user in mmWave channel, which can approach the rate limit achieved by unconstrained digital beamforming solutions. However, most researches about hybrid structure focus on transmitter not receiver. They always investigate different precoding methods for specific proposes. It should be pointed out that since the hybrid structure is used at the receiver and the application target is different in this paper, the derivation of many parameters is entirely different from the current studies.

The remainder of this paper is organized as follows. Section \textrm{II} describes the system model. In Section \textrm{III},  a robust hybrid ADB scheme of combining NSP and DL is proposed to combat the  DOA measurement errors and dramatically reduce the circuit cost is proposed. Section \textrm{IV} presents simulation results to evaluate the performance of our proposed algorithm. Finally, our conclusions are drawn in Section \textrm{V}.

Notation: throughout the paper, matrices, vectors, and scalars are denoted by letters of bold upper case, bold lower case, and lower case, respectively. Signs $(\cdot)^T$ and $(\cdot)^H$ denote transpose and conjugate transpose,~respectively. Notation $\mathbb{E}\{\cdot\}$ stands for the expectation operation. Matrices $\textbf{0}_M\times{N}$ denotes the $M\times{N}$ matrix of all zeros.

\section{System Model}
In this paper, we consider a partially connected receive structure, where each antenna is connected to one phase shifter.
\begin{figure}[h]\label{Sys_mod}
\centering
\includegraphics[width=0.5\textwidth]{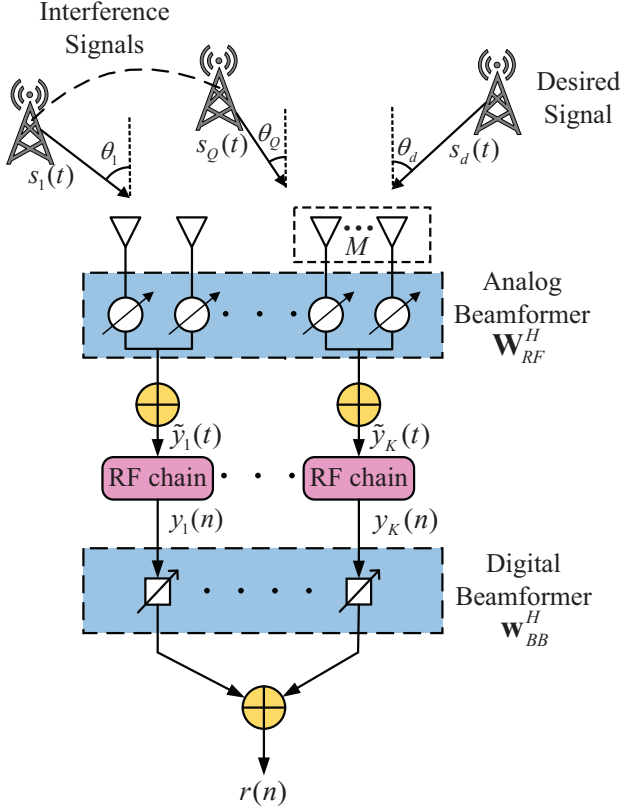}\\
\caption{The uniform linear array (ULA) hybrid beamforming structure with sub-connected architecture.}\label{receiver}
\end{figure}
In Fig.~1, one desired emitter transmits the signal $s_d(t)e^{j2\pi{f_c}t}$, where $s_d(t)$ is the baseband signal of our desired signal, and
$Q$ interference emitters transmit the narrow band signals: $s_1(t)e^{j2\pi{f_c}t}, \ldots, s_Q(t)e^{j2\pi{f_c}t}$, where $s_1(t), \ldots, s_Q(t)$ are the baseband signals of $Q$ interference signals.
These $Q+1$ signals are transmitted on the same frequency band, and then are incident on the hybrid receive array as shown in Fig.~1.
Hence, the $Q$ interference signals can be seen as the co-channel interference (CCI) for the receiver.

We assume that the linear uniform linear array (ULA) is used at the receiver, which consists of $N$ omnidirectional antenna elements and is equally divided into $K$ disjoint subsets, i.e., each subset has $M=N/K$ antennas.
It is further assumed that the desired signal comes from the DOA $\theta_d$, while the $Q$ interference signals come from the DOAs: $\theta_1, \ldots, \theta_Q$, respectively.

Then, for $k=1,2,\ldots,K$, the $k$th subarray output $\tilde{y}_k(t)$ can be represented as
\begin{align}\label{y_k_tilde}
\tilde{y}_k(t)&=\sum_{m=1}^{M}s_d(t)e^{(2\pi{f_c}t-2\pi{f_c}\tau_d-\alpha_{k,m})} \notag \\
&+\sum_{q=1}^{Q}\sum_{m=1}^{M}s_q(t)e^{(2\pi{f_c}t-2\pi{f_c}\tau_q-\alpha_{k,m})}+n_{k}(t).
\end{align}
Here, $\tau_i$ denotes the propagation delay of the received signal with the DOA $\theta_i$ for $i=d,1,2,\ldots,Q$, and can be given by
\begin{align}\label{tau_d}
\tau_i=\tau_0-\frac{\left((k-1)M+m-1\right)d}{c}\sin{\theta_i},
\end{align}
where $\tau_0$ is the propagation delay from the emitter to the first element on the array, $c$ is the speed of light, and $d$ denotes the antenna spacing.
In (\ref{y_k_tilde}), $\alpha_{k,m}$ is the corresponding phase for analog beamformer $\mathbf{W}_{RF}$ corresponding to $m$th antenna of subarray $k$. Stacking all $K$ subarray outputs in (\ref{y_k_tilde}) forms the following matrix-vector notation
\begin{align}\label{vector_y_tilde}
\tilde{\mathbf{y}}(t)=e^{j2\pi{f_c}t}\mathbf{W}_{RF}^{H}\mathbf{A}\mathbf{s}(t)+\mathbf{n}(t),
\end{align}
where $\mathbf{s}(t)=[s_d(t), s_1(t), \cdots, s_Q(t)]^T$,  and  $\mathbf{n}(t)=[n_1(t), n_2(t), \cdots, n_K(t)]^{T}$ is an additive white Gaussian noise (AWGN) with each element being Gaussian distribution $\mathcal{CN}(0, \sigma_n^2)$, whose entries are independent identically distributed, and the steering matrix $\mathbf{A}$ is defined by
\begin{align}
\mathbf{A}=\left[\mathbf{a}(\theta_d), \mathbf{a}(\theta_1), \cdots, \mathbf{a}(\theta_Q)\right],
\end{align}
where $\mathbf{a}(\theta)$ is the so-called array manifold
\begin{align}
\mathbf{a}(\theta)=\left[1, e^{j\frac{2\pi}{\lambda}d\sin{\theta}}, \cdots, e^{j\frac{2\pi}{\lambda}(N-1)d\sin{\theta}}\right]^T,
\end{align}
 and the $\mathbf{W}_{RF}$ is an ${N}\times{K}$ phase shift matrix as follows
 \begin{align}\label{W_RF}
\mathbf{W}_{RF}
=\begin{bmatrix}
\mathbf{f}_{1} & \mathbf{0} & \cdots & \mathbf{0}\\
\mathbf{0} & \mathbf{f}_{2} & \cdots & \mathbf{0}\\
\vdots & \vdots & \ddots & \vdots\\
\mathbf{0} & \mathbf{0} & \cdots & \mathbf{f}_{K}\\
\end{bmatrix},
\end{align}
where $\mathbf{f}_{k}=\frac{1}{\sqrt{M}}\left[e^{j\alpha_{1,k}},e^{j\alpha_{2,k}},\cdots,e^{j\alpha_{M,k}}\right]^{T}$ is the analog beamforming vector of the $k$th subarray. The RF signal vector $\tilde{\mathbf{y}}(t)$ in (\ref{vector_y_tilde}) passes through $K$ parallel RF chains, containing the corresponding down converters and ADCs. Thus, we have the following baseband signal vector
\begin{align}
\mathbf{y}(l)=\mathbf{W}_{RF}^{H}\mathbf{A}\mathbf{s}(l)+\mathbf{n}(l).
\end{align}
Via digital beamforming operation, the above signal vector becomes
\begin{align}
r(l)=\mathbf{w}_{BB}^{H}\mathbf{W}_{RF}^{H}\mathbf{A}\mathbf{s}(l)+\mathbf{w}_{BB}^{H}\mathbf{n}(l).
\end{align}
where  $\mathbf{w}_{BB}=[w_1, w_2, \cdots, w_K]^T$ stands for the digital beamformer.

\section{Proposed Robust Hybrid ADB Scheme}
In practical applications, we can only obtain the estimated value of DOA. If we have the prior knowledge of DOA measurement errors such as their statistical knowledge, a robust hybrid ADB scheme is proposed  and designed to achieve an obvious interference reduction and combat the effect of performance loss produced by DOA measurement errors.

\subsection{Design of Total Beamforming Vector}\label{sec_A}
The total steering matrix $\mathbf{A}$ is expressed as $\left[\mathbf{a}(\theta_d)~~\mathbf{A}_I\right]$, where $\mathbf{A}_I$ is made up of all steering vectors of interference signals.
The total beamforming vector is defined by $\mathbf{W}_{RF}\mathbf{w}_{BB}=\mathbf{v}$ and can be viewed as one single optimization variable.
According to the criterion of null space projection (NSP), we consider the following optimization problem:
\begin{align}\label{null_space}
&\mathrm{maximize}~~\|\mathbf{v}^{H}\mathbf{a}(\theta_d)\|^2\\\nonumber
&\mathrm{subject~to}~~\mathbf{A}_{I}^{H}\mathbf{v}=\mathbf{0}.
\end{align}

In the presence of DOA measurement errors,  the ideal DOA can be represented as
\begin{align}\label{DOA_mea_err}
\theta=\hat{\theta}+\Delta{\theta},
\end{align}
where $\hat{\theta}$ denotes the estimated DOA, and $\Delta\theta$ denotes the measurement error of direction angle.
Then, by using (\ref{DOA_mea_err}), the optimization problem~\eqref{null_space} can be rewritten as
\begin{align}\label{robust_null_space}
&\mathrm{maximize}~~\|\mathbf{v}^{H}\mathbf{a}(\hat{\theta}_d+\Delta\theta_d)\|^2\\\nonumber
&\mathrm{subject~to}~~\mathbf{A}_{I}^{H}(\Theta_I)\mathbf{v}=\mathbf{0},\\\nonumber
&~~~~~~~~~~~~~~~\Theta_I=\{\hat{\theta}_1+\Delta\theta_1,\cdots,\hat{\theta}_Q+\Delta\theta_Q\},
\end{align}

By assuming that $\Delta\theta$ is uniformly distributed over the interval $[-\varepsilon, \varepsilon]$, we define the probability distribution of $\Delta\theta$, denoted by $p(\Delta\theta)$, as
\begin{equation}\label{pdf}
   p\left(\Delta\theta\right)=\left\{
   \begin{aligned}
    &\frac{1}{2\varepsilon}, &-\varepsilon\leq\Delta\theta\leq\varepsilon,\\
    &0,                       &\mathrm{otherwise},
   \end{aligned}
   \right.
\end{equation}
where $\varepsilon$ is the maximum DOA estimation error.
Due to the effect of DOA estimation error $\Delta\theta$, the exact DOA $\theta$ can also be viewed as a uniform distribution with nonzero mean.
As such, we have
\begin{align}
\mathbb{E}[\mathbf{a}(\theta_d)]=\mathbb{E}[\mathbf{a}(\hat{\theta}_d+\Delta\theta_d)]\triangleq\mathbf{r}.
\end{align}
The $i$th element of $\mathbf{r}$ is
\begin{align}\label{r_i}
\mathbf{r}_i=&\int_{-\varepsilon}^{\varepsilon}e^{j\frac{2\pi}{\lambda}(i-1)d\sin{(\hat{\theta}_d+\Delta\theta_d)}}\times{p\left(\Delta\theta\right)}\mathrm{d}(\Delta\theta_d)\notag\\
=&\int_{-\varepsilon}^{\varepsilon}e^{j\frac{2\pi}{\lambda}(i-1)d(\sin{\hat{\theta}_d\cos\Delta\theta_d}+\cos{\hat{\theta}_d\sin\Delta\theta_d})} \notag \\
&\times{p\left(\Delta\theta\right)}\mathrm{d}(\Delta\theta_d) \notag \\
=&\frac{1}{2\pi}\int_{-\pi}^{\pi}e^{a_i\cos{cx}+b_i\sin{cx}}\mathrm{d}x
\end{align}
where $a_i\triangleq{j}\frac{2\pi}{\lambda}(i-1)d\sin{\hat{\theta_d}}$, $b_i\triangleq{j}\frac{2\pi}{\lambda}(i-1)d\cos{\hat{\theta_d}}$ and $c\triangleq{\frac{\varepsilon}{\pi}}$.

Furthermore, we have
\begin{align}
\mathbb{E}[\mathbf{A}_{I}^{H}(\Theta_I)]\triangleq{\mathbf{R}},
\end{align}
The entry of the $i$th row and the $q$th column of $\mathbf{R}$ is
\begin{align}
\mathbf{R}_{iq}=\frac{1}{2\pi}\int_{-\pi}^{\pi}e^{a_{iq}\cos{cx}+b_{iq}\sin{cx}}\mathrm{d}x,
\end{align}
where $a_{iq}\triangleq{j}\frac{2\pi}{\lambda}(i-1)d\sin{\hat{\theta}_q}$, $b_{iq}\triangleq{j}\frac{2\pi}{\lambda}(i-1)d\cos{\hat{\theta}_q}$, and $c\triangleq{\frac{\varepsilon}{\pi}}$.

Substituting $\mathbf{r}$ and $\mathbf{R}$ into (\ref{robust_null_space}) forms the following robust optimization problem
\begin{align}\label{robust_null_space-V1}
&\mathrm{maximize}~~\|\mathbf{v}^{H}\mathbf{r}\|^2\\\nonumber
&\mathrm{subject~to}~~\mathbf{R}^{H}\mathbf{v}=\mathbf{0}.
\end{align}

In accordance with the above equality constraint,  the total beamforming vector $\mathbf{v}$ is orthogonal  to the null space of $\mathbf{R}$.  To construct $\mathbf{v}$,  the singular value decomposition (SVD) operation is performed on conjugate transpose of interference matrix $\mathbf{R}$  as follows: $\mathbf{R}=\mathbf{U}\sum\mathbf{V}^{H}$, where $\mathbf{U}$ and $\mathbf{V}$ are the unitary matrices with $\mathbf{U}\in\mathbb{C}^{Q\times{Q}}$, and $\mathbf{V}\in\mathbb{C}^{N\times{N}}$.
Here, $\sum\in\mathbb{C}^{Q\times{N}}$ is a rectangle matrix with singular values on its main diagonal and all off-diagonal elements being zeros, i.e. $\sum=\left[\text{diag}\left\{\sigma_1^2,\sigma_2^2,\cdots,\sigma_{Q}^2\right\},\mathbf{0}_{Q\times(N-Q)}\right]$.
According to the equality constraint in (\ref{robust_null_space-V1}),  $\mathbf{v}$ can be given by
a linear combination of the $N-Q$ most right column vectors of matrix $\mathbf{V}$, i.e.,
\begin{align}
\mathbf{v}=\mathbf{F}\tilde{\mathbf{v}},
\end{align}
where $\mathbf{F}$ is the $N-Q$ most right columns of $\mathbf{V}$, and $\tilde{\mathbf{v}}$ is a column vector with each entry controlling the linear combination of right singular vectors and it has normalized power $\mathbb{E}(\tilde{\mathbf{v}}^H\tilde{\mathbf{v}})=1$.

Therefore, the optimization problem in (\ref{null_space}) can be rewritten as
\begin{align}
\mathrm{maximize}~~\|\tilde{\mathbf{v}}^{H}\mathbf{F}^{H}\mathbf{r}\|^2,
\end{align}
which directly yields
\begin{align}
\tilde{\mathbf{v}}_{opt}=\frac{\mathbf{F}^{H}\mathbf{r}}{\|\mathbf{F}^{H}\mathbf{r}\|},
\end{align}
and
\begin{align}\label{v_opt}
\mathbf{v}_{opt}=\mathbf{F}\tilde{\mathbf{v}}_{opt}.
\end{align}

This completes the design of the total beamforming vector.

\subsection{DL-based Digital Beamformer}
In what follows, given the initial value of analog beamformer $\mathbf{W}_{RF_0}$, we show how to optimize the digital beamforming vector $\mathbf{w}_{BB}$.
A wise choice to design $\mathbf{W}_{RF_0}$ is to make the array point towards the DOA of the desired signal at first, i.e.,
\begin{align}
\alpha_{k,m,0}=\frac{2\pi}{\lambda}\left((k-1)M+m-1\right)d\sin({\hat{\theta}_d}+\Delta\theta_d).
\end{align}
where $\alpha_{k,m,0}$ is the initial corresponding phase of $\mathbf{W}_{RF_0}$.

Similar to the derivation of $\mathbf{r}$ and $\mathbf{R}$, in order to calculate the initialization value $\mathbf{W}_{RF_0}$, the expectation of $\sin({\hat{\theta}_d}+\Delta\theta_d)$ is adopted to replace the exact $\sin{\theta_d}$, i.e.,
\begin{align}
\mathbb{E}[\sin{\theta_d}]&=\mathbb{E}[\sin({\hat{\theta}_d+\Delta\theta_d})]\\\nonumber
&=\int_{-\varepsilon}^{\varepsilon}\sin({\hat{\theta}_d+\Delta\theta_d})\times{p(\Delta\theta_d)}\mathrm{d}(\Delta\theta_d)\\\nonumber
&=\frac{1}{\varepsilon}\sin{\hat{\theta}_d}\sin{\varepsilon}.
\end{align}
Thus, the corresponding phase of $\mathbf{W}_{RF_0}$ is given by
\begin{align}\label{robust_phase}
\alpha_{k,m,0}=\frac{2\pi}{\varepsilon\lambda}\left((k-1)M+m-1\right)d\sin{\hat{\theta}_d}\sin{\varepsilon}.
\end{align}

The steering vector of subarray $\mathbf{a}_{sub}$ is generated with known $\mathbf{W}_{RF_0}$. Let us define the subarray steering vector of the desired signal as follows
\begin{align}\label{W_RF0}
\mathbf{a}_{sub}(\hat{\theta}_d+\Delta\theta_d)=\mathbf{W}_{RF_0}^{H}\mathbf{a}(\hat{\theta}_d+\Delta\theta_d).
\end{align}

Due to that the perfect information of noise and signals is usually unavailable in practice, the sampling covariance matrix is adopted instead. Since the Capon beamforming method \cite{Capon} is sensitive to modeling errors, it is not robust for modeling mismatch, which will deteriorate the output of Capon beamformer seriously. Therefore, instead of Capon method, the  DL method is employed to design digital beamforming vector, which is robust to model mismatch. Via the regularization operation on Capon method, the optimization problem of using DL method to optimize the beamforming vector $\mathbf{w}_{BB}$ can be casted as
\begin{align}\label{w_BB_problem}
\mathop{\mathrm{minimize}}\limits_{\mathbf{w}_{BB}}~~~~&{\mathbf{w}_{BB}^{H}\left(\mathbf{\hat{R}}+\gamma\mathbf{I}\right)\mathbf{w}_{BB}}\\\nonumber
\mathrm{subject~to}~~~&\mathbf{w}_{BB}^{H}\mathbf{a}_{sub}(\hat{\theta}_d+\Delta\theta_d)=1,
\end{align}
where $\gamma$ denotes the DL factor, and $\mathbf{\hat{R}}$ is  the sampling covariance matrix corresponding to $K$ subarrays,
\begin{align}
\mathbf{\hat{R}}=\frac{1}{L}\sum_{l=1}^{L}\mathbf{y}^{H}(l)\mathbf{y}(l).
\end{align}
where $L$ is the number of snapshots. Applying the Lagrange multiplier method to the optimization problem (\ref{w_BB_problem}), the associated Lagrangian  function has the following form
\begin{align}
f\left(\mathbf{w}_{BB},\lambda\right)&=\mathbf{w}_{BB}^{H}\left(\mathbf{\hat{R}}+\gamma\mathbf{I}\right)\mathbf{w}_{BB}\\\nonumber
&+\lambda\left(\mathbf{w}_{BB}^{H}\mathbf{a}_{sub}(\hat{\theta}_d+\Delta\theta_d)-1\right),
\end{align}
where the scalar $\lambda$ is the lagrange multiplier. By taking the derivation of $f\left(\mathbf{w}_{BB},\lambda\right)$ with respect to $\mathbf{w}_{BB}$ and letting it equal zero, the digital beamformer is
\begin{align}\label{w_BB_DL_lambda}
\mathbf{w}_{BB}=-\lambda\left(\mathbf{\hat{R}}+\gamma\mathbf{I}\right)^{-1}\mathbf{a}_{sub}(\hat{\theta}_d+\Delta\theta_d).
\end{align}
Substitute (\ref{w_BB_DL_lambda}) into the equation constraint of (\ref{w_BB_problem}), then $\mathbf{w}_{BB}$ is expressed as
\begin{align}
\mathbf{w}_{BB}=\frac{\left(\mathbf{\hat{R}}+\gamma\mathbf{I}\right)^{-1}\mathbf{W}_{RF_0}^{H}\mathbf{a}(\hat{\theta}_d+\Delta\theta_d)}{\mathbf{W}_{RF_0}^{H}\mathbf{a}(\hat{\theta}_d+\Delta\theta_d)\left(\mathbf{\hat{R}}+\gamma\mathbf{I}\right)^{-1}\mathbf{W}_{RF_0}^{H}\mathbf{a}(\hat{\theta}_d+\Delta\theta_d)}.
\end{align}
Similar to the derivation of total beamformer, $\mathbf{r}$ is used to replace $\mathbf{a}(\hat{\theta}_d+\Delta\theta_d)$, therefore, the above equation is written as
\begin{align}\label{robust_wbb}
\mathbf{w}_{BB}=\frac{\left(\mathbf{\hat{R}}+\gamma\mathbf{I}\right)^{-1}\mathbf{W}_{RF_0}^{H}\mathbf{r}}{\mathbf{W}_{RF_0}^{H}\mathbf{r}\left(\mathbf{\hat{R}}+\gamma\mathbf{I}\right)^{-1}\mathbf{W}_{RF_0}^{H}\mathbf{r}}.
\end{align}
Observing the above expression, the main advantage of the DL method  ensures that the diagonal loading matrix $\mathbf{\hat{R}}+\gamma\mathbf{I}$ is invertible by adding white noise to diagonal elements of sampling covariance matrix $\mathbf{\hat{R}}$.

\subsection{Analytic Analog Beamformer}
Now, we turn to the construction of the analog beamformer $\mathbf{W}_{RF}$. To simultaneously reduce the interference and maximize the received power of desired signal, we model the problem of optimizing $\mathbf{W}_{RF}$ as follows
\begin{align}
\mathrm{minimize}~~~~&\|\mathbf{v}_{opt}-\mathbf{W}_{RF}\mathbf{w}_{BB}\|,
\end{align}

Based on (\ref{W_RF}) and expression of $\mathbf{f}_{k}$, the above unconstrained optimization problem can be decomposed into the following $N$ independent sub-optimization problems:
\begin{align}
\mathrm{minimize}~~~~\|\mathbf{v}_{{opt}_{(k-1)\times{M}+m}}-\frac{1}{\sqrt{M}}e^{j\alpha_{k,m}}{w}_{k}\|,
\end{align}
where $k\in\{1,2,\cdots,K\}$, $m\in\{1,2,\cdots,M\}$, and $\mathbf{v}_{{opt}_{k}}$ and ${w}_{k}$  represent the $k$th element of $\mathbf{v}_{opt}$ and $\mathbf{w}_{BB}$, respectively. The above sub-optimization directly yields the following closed-form solution
\begin{align}\label{phase_k,m}
\alpha_{k,m}=\angle\left(\frac{\mathbf{v}_{{opt}_{(k-1)\times{M}+m}}}{{w}_{k}}\right).
\end{align}
Based on the above construction, we summarize our robust hybrid ADB algorithm in Algorithm I.
\begin{algorithm}[h]
\caption{The proposed robust hybrid ADB algorithm} 
\hspace*{0.02in} {\bf Input:} $\mathbf{r}$, $\mathbf{R}$
\begin{algorithmic}[1]
\State Initialize $\mathbf{W}_{{RF}_0}$ by (\ref{robust_phase});
\State Calculate $\mathbf{v}_{opt}$ based on $\mathbf{r}$ and $\mathbf{R}$;
\State  DL-based digital beamforming vector $\mathbf{w}_{BB}$ is constructed
in accordance with (\ref{robust_wbb});
\State Reconstruct $\alpha_{k,m}$ by (\ref{phase_k,m});
\end{algorithmic}
\hspace*{0.02in} {\bf Output:} $\mathbf{W}_{RF}$,$\mathbf{w}_{BB}$
\end{algorithm}

\section{Simulation Results}
In this section, we present simulation results to examine the performance of the proposed robust hybrid ADB algorithm.
We consider $N=32$,  $K=4$, and antenna spacing $d$ =$0.5\lambda$.
It is assumed that the desired DOA $\theta_d=60^\circ$, and the DOAs of two interference sources are $30^\circ$ and $-15^\circ$, respectively.


\begin{figure}[h]
\centering
\includegraphics[width=0.5\textwidth]{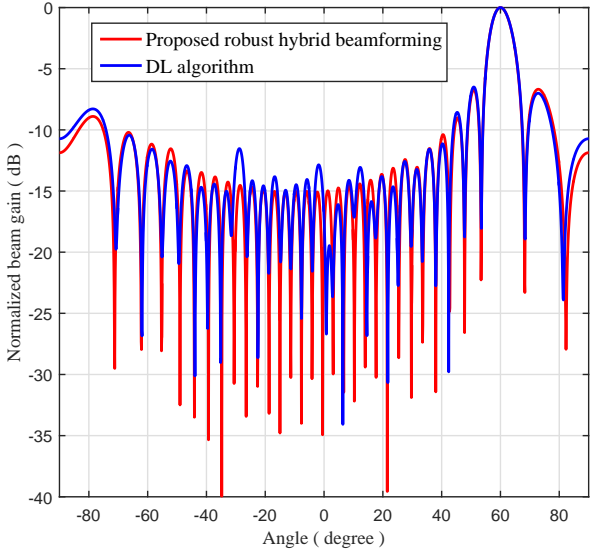}\\
\caption{Normalized beam gain patterns for DL algorithm and proposed robust hybrid ADB without DOA measurement errors.}\label{beam}
\end{figure}
First, we consider the situation with perfect DOA knowledge.
Fig.~\ref{beam} shows that the curves of beamforming gain versus direction of the DL method and the proposed algorithm.
It can be seen that the proposed method can attenuate the beam gain by at least $30dB$ in the direction of interference signals, where signal to noise ratios (SNR) of desired and interference sources are $0dB$ and $15dB$, respectively.
In contrast with the DL algorithm, our method performs well along the direction of interference and has the same gain in the desired direction.
\begin{figure}[h]
\centering
\includegraphics[width=0.5\textwidth]{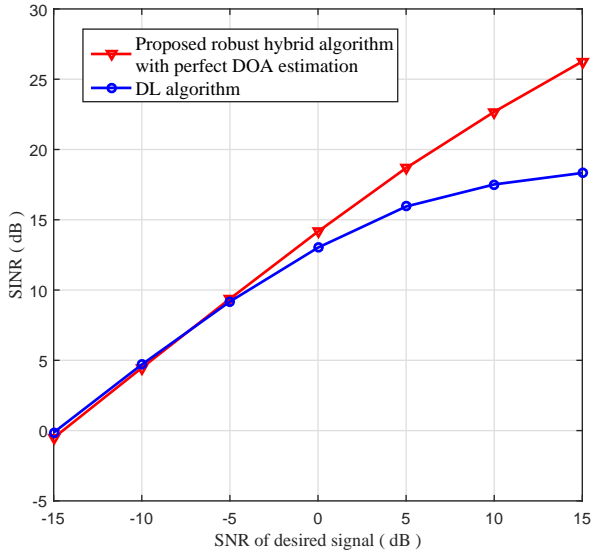}\\
\caption{Curves of SINR versus SNR for DL and proposed robust hybrid ADB methodS without DOA measurement errors.}\label{sinr_exact}
\end{figure}

In Fig.~\ref{sinr_exact}, we present the curves of signal to interference plus noise ratio (SINR) versus SNR of the desired signal by ranging SNR from -15dB to 15dB.
It is seen that the proposed hybrid beamformer almost has the same performance as DL method in the low SNR region.
Moreover, as SNR increases, our algorithm has a stronger ability to reduce interference. Specifically, when the SNR of the desired signal is less than -5dB, the SINR curve of our proposed algorithm and DL method are identical.
Nevertheless, when SNR is higher than -5dB, the SINR curve of the proposed algorithm is above that of DL.

According to the estimator presented by \cite{J.Sheinvald}, the root mean square (RMS) is less than $1^\circ$ under the condition that the SNR of estimated source is 0dB. In order to examine the robust performance of our proposed method in extreme conditions, it is  supposed that $\Delta\theta$ is uniformly distributed and there exists a maximum estimation error $3^\circ$, i.e., $\varepsilon=3^\circ$.
\begin{figure}[h]
\centering
\includegraphics[width=0.5\textwidth]{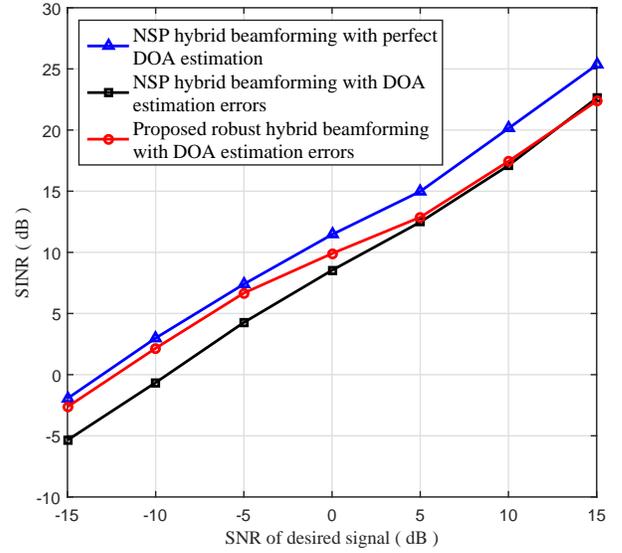}\\
\caption{Curves of SINR versus SNR with DOA estimation errors.}\label{sinr_robust}
\end{figure}

Fig.~\ref{sinr_robust} illustrates the robustness performance of our proposed algorithm. The curve of the proposed NSP algorithm with perfect DOA estimation is used as a reference. Observing this figure, we find, with increase in the value of SNR, the SINR performance of the NSP hybrid beamforming with DOA estimation errors is always almost 3dB worse than that with perfect DOA estimation. It is noted that when the desired SNR is low, considering the uniformly distributed DOA estimation errors, the robust algorithm is closer to the NSP algorithm with perfect DOA, that is, our robust hybrid beamforming can efficiently reduce interference and retain robust when there exists direction-finding errors in the low SNR region, which is the typical situations in interference reduction.
\begin{figure}[h]
\centering
\includegraphics[width=0.5\textwidth]{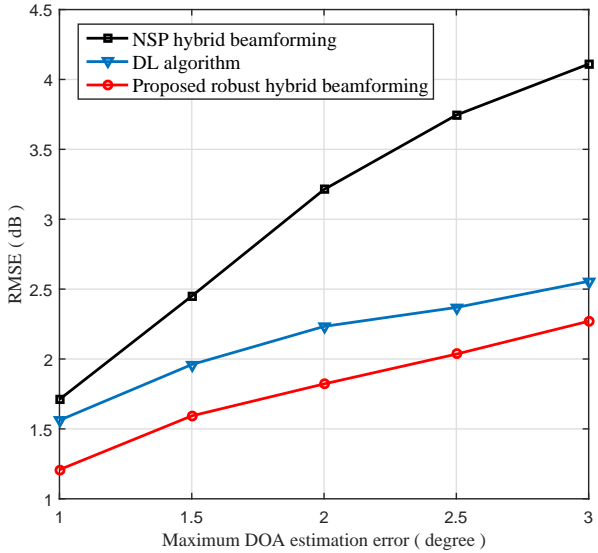}\\
\caption{Curves of RMSE versus  DOA estimation error for NSP beamforming without DOA estimation error and three methods with DOA measurement errors.}\label{delta_max}
\end{figure}

Fig.~\ref{delta_max} plots  the curves of RMSE versus maximum measurement angle errors $\varepsilon$ for the proposed hybrid ADB while the NSP with perfect DOA is used as a reference. By our calculation, when there exist DOA estimation errors, the  RMSE differences between the reference and three methods grow gradually with angle error. It is noted that, with the increase of $\varepsilon$, the RMSEs of NSP hybrid beamforming, DL method and the robust hybrid beamforming become worse. However, the RMSE of robust hybrid beamforming is always lower than that of DL algorithm and NSP hybrid beamforming. This  means that it has a better performance.
\begin{figure}[h]
\centering
\includegraphics[width=0.5\textwidth]{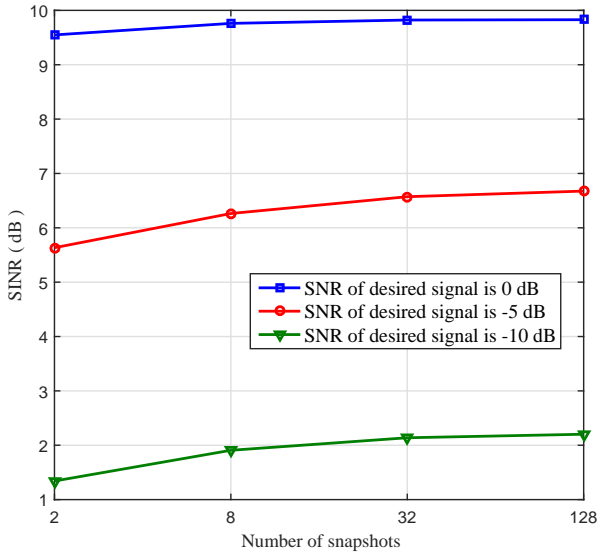}\\
\caption{Curves of SINR versus snapshot number for three different SNRs}\label{snapshot}
\end{figure}

Finally,  Fig.~\ref{snapshot} illustrates the impact of snapshot number on the robust hybrid ADB scheme proposed. From  Fig.~\ref{snapshot}, it is seen that the performance of the proposed robust hybrid beamforming is gradually improved with increasing the snapshot number. When the receive SNR of the desired signal is in the low SNR region, the snapshot number has bigger impact on the performance improvement. Once the number of snapshots reaches the limit value $32$, further increasing it has a trivial impact on performance improvement.
\section{Conclusion}
 In this paper, a robust hybrid ADB scheme, has been proposed, which are based on DL method, where the NSP rule is adopted to design the total beamforming vector. In the case of perfect DOA available, the proposed scheme achieves a substantial beam gain over DL algorithm in the interference direction. In the presence of DOA measurement errors, the proposed robust hybrid ADB, with uniformly distributed angle errors, shows a good robustness compared to DL algorithm and the NSP hybrid method. As the maximum angle error becomes larger, its SINR performance over non-robust schemes such as DL  becomes more significant. In the coming future, the proposed method may be potentially applied to future directional modulation networks, satellite communcations, mmWave communications, and unmanned aerial vehicle (UAV).

\ifCLASSOPTIONcaptionsoff
  \newpage
\fi

\bibliographystyle{IEEEtran}
\bibliography{IEEEfull,cite}

\end{document}